\documentclass{IAU-TrA}
\usepackage{graphicx}

\title[COLLISION PROCESSES]
{}

\author[DIVISION~XII / COMMISSION~14 / WORKING GROUP ]
{}

\pubyear{2012}
\volume{Volume XXVIIIA}
\jname{Transactions IAU, Volume XXVIIIA}
\editors{Ian Corbett, ed.}
\begin{document}

\maketitle

{\bf

\large
\noindent
DIVISION XII / COMMISSION~14 / WORKING GROUP                       \\
COLLISION PROCESSES \\

\normalsize

\begin{tabbing}
\hspace*{45mm}  \=                                                   \kill
CO-CHAIRS       \> Gillian Peach                                     \\
                \> Milan S. Dimitrijevic                             \\

\end{tabbing}

\noindent
TRIENNIAL REPORT 2009-2012
}

\firstsection % if your document starts with a section,
              % remove some space above using this command.

\section{Introduction}

Research in atomic and molecular collision processes and spectral line
broadening has been very active since our last report (\cite{Peds09}).
Given the large volume of the published literature and the limited
space available, we have attempted to identify work most relevant to
astrophysics.  Since our report can not be
comprehensive, additional publications can be found in the databases
at the web addresses listed in the final section.  Elastic and 
inelastic collisions among electrons, atoms, ions, and molecules are
included and charge transfer can be very important in collisions between
heavy particles.

Numerous meetings on collision processes and line broadening have been
held throughout the report period.  Important international meetings that
provide additional sources of data through their proceedings are:
the 19th {\it International Conference on Spectral Line Shapes (ICSLS)}
(\cite{Gigo08}), the 7$^{th}$ {\it Serbian Conference on Spectral Line Shapes
in Astrophysics} (SCSLSA) (\cite{Popo09}), the XXVI {\it International
Conference on Photonic, Electronic, and Atomic Collisions} (ICPEAC)
(\cite{Orel09}), the 20$^{th}$ {\it ICSLS} (\cite{Lewi10}), the 22$^{nd}$
{\it International Conference on Atomic Physics} (ICAP) (\cite{Bach11}), 
and the 7$^{th}$ {\it International Conference on Atomic and Molecular
Data and their Applications} (\cite{Bern11}).  The 8$^{th}$ SCSLSA and the
XXVII$^{th}$ ICPEAC took place in June and July 2011 and 
their proceedings will
be published in {\it Baltic Astronomy} and {\it Journal of Physics:
Conference Series}, respectively.

\section{Electron collisions with atoms and molecules}

Collisions of electrons with atoms, molecules and atomic and molecular ions
are the major excitation mechanism for a wide range of astrophysical
environments. In addition, electron collisions play an important role in
ionization and recombination, contribute to cooling and heating of the gas,
and may contribute to molecular fragmentation and formation.  In the
following sections we summarize recent work on collisions for astrophysically
relevant species, including elastic scattering, excitation, ionization,
dissociation, recombination and electron attachment and detachment.

A review has been published of the atomic data necessary for the non-LTE
analysis of stellar spectra (\cite{Mash09}).  Other references are listed
below for scattering by the atoms and molecules specified.

\subsection{Electron scattering by neutral atoms}

Elastic scattering:
H (screened Coulomb interactions) (\cite{Zhan10}),
Mg (\cite{Zats09}), 
Ar (\cite{Garg08}),
I (\cite{Zats11}),
Mn, Cu, Zn, Ni, Ag, Cd (\cite{Felf11}),
Rb, Cs, Fr (\cite{Gang10}).
\\[1ex]
Excitation:
H (screened Coulomb interactions) (\cite{Zhan10,Zhan11}),
He(2$^{1,3}$S) (\cite{Wang09,Wang10}),
He, Ne (\cite{Kret08}),
Mg (\cite{Zats09}),
Ar (\cite{Garg08}).
\\[1ex]
Ionization: He (\cite{Bray10,Renx11}),
He(2$^1$S) (\cite{Wang10}),
Ar (\cite{Garg08}).
\\[1ex]
Total cross section: Na (\cite{Jiao10}).

\subsection{Electron scattering by atomic ions}

Elastic scattering: Mg$^+$, Ca$^+$ (\cite{Mitr08}).
\\[1ex]
Excitation:
Hydrogen isoelectronic sequence  Cr$^{23+}$--Ni$^{27+}$
  (\cite{Male11}), 
Lithium isoelectronic sequence Be$^+$--Kr$^{33+}$
  (\cite{Lian11a}),
Ne$^{3+}$, Ne$^{6+}$ (\cite{Ludl11}), 
Neon isoelectronic sequence Na$^+$--Kr$^{26+}$ (\cite{Lian10}),
Sodium isoelectronic sequence Mg$^+$--Kr$^{25+}$ (\cite{Lian09b}),
Mg$^{4+}$ (\cite{Huds09}), Mg$^{8+}$ (\cite{Delz08}),
Si$^{9+}$ (\cite{Lian09a}),
S$^{8+}$--S$^{11+}$ (\cite{Lian11b}),
K$^+$ (\cite{Taya10}),
Fe$^{10+}$ (\cite{Delz10}),
Fe$^{12+}$ (\cite{Stor10}),
Fe$^{18+}$ (\cite{Butl08}), 
Ni$^{10+}$ (\cite{Agga08a}),
Ni$^{18+}$ (\cite{Agga08c}).
\\[1ex]
Recombination:
H$^+$ (\cite{Chlu10}),
N$^+$ (\cite{Fang11}),
Aluminium isoelectronic sequence Si$^+$--Zn$^{17+}$ (\cite{Abde11}),
Argon isoelectronic sequence K$^+$--Zn$^{12+}$ (\cite{Niko10}),
Fe$^7+$, Fe$^{8+}$ (\cite{Schm08}),
Selenium ions Se$^{q+}$, $q = 1-6$ (\cite{Ster11}).
\\[1ex]
Energy levels, radiative and excitation rates:
O$^{3+}$ (\cite{Agga08b,Keen09}),
O$^{6+}$ (\cite{Agga08d}),
Si$^{+}$ (\cite{Baut09}),
Ar$^{17+}$ (\cite{Agga08}),
Selenium ions Se$^{q+}$, $q = 1-6$ (\cite{Ster11}).
\\[1ex]
X-ray line emission: Na$^{9+}$ (\cite{Phil10}).\\
Radiative and Auger decay: Aluminium ions Al$^{q+}$,
$q = 0-11$ (\cite{Palm11}).

\subsection{Electron scattering by molecules}

Elastic scattering:
H$_2$O (\cite{Liuj10}),
(H$_2$O)$_2$ (\cite{Bouc08}),
Li$_2$ (\cite{Tara08}),
CO (\cite{Alln10}),
NH (\cite{Rajv10}), 
NO$_2$ (\cite{Munj09}),
S$_2$ (\cite{Rajv11}),
SO$_2$ (\cite{Mach11}),
SOS (\cite{Kaur10}),
S$_3$ (\cite{Kaur11}).\\
Electron exchange:
O$_2$, NO, NO$_2$ (\cite{Holt09}).
\\[1ex]
Excitation:
H$_2$ (\cite{Kret08}),
Li$_2$ (\cite{Tara08}),
CO (\cite{Alln10}),
N$_2$ (\cite{Kato10,John10,Mava11}),
NH (\cite{Rajv10}),
NO$_2$ (\cite{Munj09}),
S$_2$ (\cite{Rajv11}),
SOS (\cite{Kaur10}),
S$_3$ (\cite{Kaur11}),
C$_2$H$_4$ (\cite{Cost08}).
\\[1ex]
Ionization:
N$_2$ (\cite{Goch10}),
NH (\cite{Rajv10}),
SiCl, SiCl$_2$, SiCl$_3$, SiCl$_4$ (\cite{Koth11}),
SO, SO$_2$ (\cite{Vino08}),
SOS (\cite{Kaur10}),
S$_2$ (\cite{Rajv11}),
S$_3$ (\cite{Kaur11}).
\\[1ex]
Total cross sections:
SO, SO$_2$, SO$_2$Cl$_2$, SO$_2$ClF, SO$_2$F$_2$ (\cite{Josh08}),
SO$_2$ (\cite{Mach11}),
C$_2$H$_4$O (\cite{Szmy08}),
CH$_3$OH, CH$_3$NH$_2$ (\cite{Vino08}).
\\[1ex]
Dissociative processes:
H$_2$ (\cite{Celi09,Bell10,Celi11}),
HCl (\cite{Fedo10}),
HCl, DCl, HBr, DBr (\cite{Fedo08}),
C$_2$H$_2$ (\cite{Chou08}).
\\[1ex]
Attachment:
C$_2$H, C$_2$N (\cite{Harr11}).

\subsection{Electron scattering by molecular ions}

Detachment: C$_2^-$ (\cite{Halm08}).\\
Excitation: C$_2^-$ (\cite{Halm08}),
  CO$^+$ (\cite{Stau09}).\\
Rotational cooling: HD$^+$ (\cite{Shaf09}).
\\[1ex]
Dissociative and recombination processes:
HD$^+$ (\cite{Fifi08,Taka09,Stro11}),
H$_3^+$ (\cite{Glos09}),
HF$^+$ (\cite{Roos08,Roos09}),
CH$_3^+$, CD$_3^+$ (\cite{Baha09}),
CD$_3$OCD$_2^+$, (CD$_3$)$_2$OD$^+$ (\cite{Hamb10a}),
CD$_3$CDOD$^+$, CH$_3$CH$_2$OH$_2^+$ (\cite{Hamb10b}).

\section{Collisions between heavy particles}

A review entitled `Energetic ion, atom and molecule reactions and
excitation in H$_2$ discharges' has recently been published
(\cite{Phel09}).  Other references are listed below for the atomic
and molecular processes specified.

\subsection{Collisions between neutral atoms and atomic ions}

Inelastic scattering: H + H (\cite{Bark11}), Na + H (\cite{Bark10}).\\
Excitation: H + H$^+$ (\cite{Wint09}).
\\[1ex]
Charge transfer processes: H, D, T + He$^{2+}$ (\cite{Stol10}),
H$^-$ + H$^+$, mutual neutralization (\cite{Sten09}),
H + H$^+$ (\cite{Wint09}),
H + Li$^{2+}$ (\cite{Manc09}),
H + B$^{5+}$, C$^{4+}$ (\cite{Barr10}),
H + C$^{3+}$, O$^{3+}$, Si$^{3+}$ (\cite{Guev11}),
H$^+$ + Li$^+$, Be$^{2+}$, B$^{3+}$, C$^{4+}$ (\cite{Sama11a}),
He + H$^+$ (\cite{Guzm09,Harr10,Fisc10}),
He + H$^+$, He$^{2+}$ (\cite{Zapu09}),
He + H$^+$, He$^+$, He$^{2+}$ (\cite{Scho09}),
He + $^3$He$^{2+}$ (\cite{Ales11}),
He + He$^{2+}$, Li$^+$, Li$^{2+}$, Li$^{3+}$, C$^{6+}$, O$^{8+}$
  (\cite{Sama11b}),
He + C$^{3+}$ (\cite{Wuyb09}),
He + N$^{3+}$ (\cite{Liux11}),
He + O$^{3+}$ (\cite{Kamb08,Wuya09}),
He, Ne, Ar, Kr, Xe + C$^{3+}$ (\cite{Sant10}),
He$^+$ + He$^+$ (\cite{Manc09}),\\
Li + H, H$^+$ (\cite{Cabr08}),
Li + H$^+$ (\cite{Liul11}),
Ne + C$^{2+}$, C$^{3+}$, O$^{2+}$, O$^{3+}$ (\cite{Ding08}),
Na + H ion-pair production (\cite{Bark10}),
Mg + H$^+$, He$^{2+}$ (\cite{Kuma11}),
Mg + Cs$^+$ (\cite{Sabi08}),
Ar + H$^+$ (\cite{Cabr09}), Ar + O$^{3+}$ (\cite{Kamb08}).
\\[1ex]
Ionization:  H + H$^+$ (\cite{Wint09}),
He + H$^+$ (\cite{Guzm09}),
He + He$^{2+}$ (\cite{Ogur11}), 
He, Ne, Ar, Kr, Xe + H$^+$ (\cite{Mira08}),
He, Ne, Ar, Kr, Xe + C$^{3+}$ (\cite{Sant10}),
Li$^+$, Na$^+$, K$^+$, Rb$^+$ + H$^+$ (\cite{Mira08}).
\\[1ex]
Detachment: C$^-$, O$^-$, F$^-$, Na$^-$, Si$^-$, S$^-$, Cl$^-$, Ge$^-$ +
He, Ne, Ar (\cite{Jalb08}),
F$^-$, Cl$^-$, Br$^-$, I$^-$ + H$^+$ (\cite{Mira08}).
\\[1ex]
Energy loss and stopping cross sections: Li + H, H$^+$
  (\cite{Cabr08}),
H, D, T + He$^{2+}$ (\cite{Cabr11}).

\subsection{Collisions between atoms and molecules}

Dissociative processes:
H$_2$O + H$^+$ (\cite{Monc09}),
H$_2$O + He$^+$ (\cite{Garc08}).
\\[1ex]
Radiative association:
H$_2$ + H$^-$ (\cite{Ayou11}).
\\[1ex]
Excitation and/or fragmentation:
CN$^-$ + H$_2$ (\cite{Agun10}),
CO$^+$ + H (\cite{Ande08}),
CO$_2^+$ + He (\cite{Gonz08}),
CO$^+$ + H, H$_2$ (\cite{Stau09}).
\\[1ex]
Charge transfer processes:
D$_2$, O$_2$, H$_2$O, CO$_2$ + O$^{3+}$ (\cite{Kamb08}),
H$_2$ + He$^{2+}$ (\cite{Khom09}),
CH$_3$ + H$^+$ (\cite{Nago08}),
CO$_2$ + He$^+$ (\cite{Liny10}),
HCl + C$^{2+}$ (\cite{Rozs11}).
\\[1ex]
Ionization and/or capture:
H$_2$O + H$^+$, He$^{2+}$, C$^{6+}$ (\cite{Ille11}),
C$_2$H$_4$ + H$^+$ (\cite{Geta10}),
N$_2$ + H$^+$ (\cite{Goch10}).

\section{Stark broadening}

Knowledge of Stark broadening parameters (line widths and shifts) for a 
large number of atomic transitions is very important for the analysis,
interpretation and modelling of stellar spectra, circumstellar conditions
and H II regions. For hot dense stars such as white dwarfs this is often
the most important broadening mechanism.

\subsection{Developments in line broadening theory}

\cite{Rosa09} have reexamined the Stark broadening of hydrogen lines in the
presence of a magnetic field and developed an impact theory for ions, valid
for low electron densities ($N_e \le 10^{14}$ cm$^{-3}$), which takes into
account the Zeeman splitting of the atomic energy levels.  \cite{Rosa10}
have also studied numerically the role of time ordering in such plasmas, by
using a simulation code that accounts for the evolution of the microscopic
electric field generated by the charged particles moving close to the atom.
\cite{Cali10} have developed a very fast method to account for the dynamical
effects of charged particles on the spectral line shape emitted by plasmas,
based on a formulation of the frequency fluctuation model. 

\textit{Ab initio} calculations of Stark broadening parameters, i.e.
calculations where the required atomic energy levels and oscillator
strengths are determined during the calculation and are not taken from other
sources, have been considered and reviewed by \cite{Benn09}.  A book has
recently been published (\cite{Gord09}) that gives a detailed account of the 
surprising discovery in the 1960's of the radio recombination lines and their
subsequent analysis.  Even now some features have still not been satisfactorily
explained.

\subsection{Isolated lines}

For isolated lines Stark broadening is dominated by collisions
with plasma electrons.  Broadening parameters have been determined
theoretically for:

One line from the 3s-3p transition array for each of the spectra Si XI,
Ti XI, Cr XIII, Cr XIV, Fe XV, Fe XVI, Ni XVIII and Fe XXIII and two
lines from the array for K VIII, Ca IX, Sc X and Ti XI (\cite{Elab11});
two lines for 3s-3p transitions for ions C IV, N V, O VI, F VII, Na IX,
Mg X, Al XI, Si XII and P XIII and one line for Ne VIII (\cite{Elab09});
the 2s-2p resonance doublets of C IV, N V, O VI, F VII and Ne VIII ions
(\cite{Elai11}).  These calculations all use a quantum mechanical approach.

For five lines of Cu I (\cite{Zmer10}) and the lines Ne I 837.8 nm
(\cite{Chri10b}) and Ar I 737.2 nm (\cite{Chri10a}), new Stark broadening
parameters are obtained using a semiclassical perturbation approach.
A semi-empirical approach, which uses a set of wave functions obtained
from Hartree-Fock relativistic calculations and includes core
polarization effects, has been applied to 58 lines of Pb IV
(\cite{Alon10}) and 171 lines of Sn III (\cite{Aloc11}).

Broadening parameters have been obtained experimentally for the
following numbers of lines:

34 Pb I (\cite{Alon08}), 25 Pb III (\cite{Alon11}), 34 Pb IV and 4 Pb V
(\cite{Bukv11}), 28 Cd III (\cite{Djei09}, \cite{Bukv09b}), 13 Si I,
15 Si II, 28 Si III and 9 Si IV (\cite{Bukv09a}), 29 (\cite{Bukv08}) and
19 (\cite{Djur11}) Ar III, 30 Kr III (\cite{Ciri11}), 12 Ne II, 8 Kr II 
and 5 Xe II (\cite{Pela10b}), 38 Xe II (\cite{Pela09a,Pela09b}),
10 Xe III (\cite{Pela09b}), 5 Au I and 26 Au II (\cite{Djen09}), 9 Sb III
(\cite{Djen08}), 15 Mn I and 10
\linebreak
Fe I (\cite{Ziel10}), 21 Fe II
(\cite{Arag11}) and C I 833.5 nm (\cite{Bart11}).

The regularities and systematic trends of Stark broadening parameters
and reasons for deviations have been investigated within the multiplets
(\cite{Pela10b,Pela09a}), along the homologous sequence of singly-ionized 
noble gases (\cite{Pela10a}), within the spectral series 
(\cite{Chri10a}) and along isoelectronic sequences (\cite{Elab09},
\cite{Elai11}).  Also the dependence of electron- and proton-impact
Stark widths on the upper-level ionization potential within different 
series of spectral lines of neutral magnesium (\cite{Tapa11}) and as a
function of charge on the atomic core (\cite{Elab11}) have been evaluated
and discussed.  This kind of trend and regularity analysis can be useful
for the prediction of Stark broadening parameters and therefore for the
spectroscopic diagnostic of astrophysical plasmas. 

\subsection{Transitions in hydrogenic and helium-like systems}

Stark-broadened line profiles of the hydrogen Brackett series have
been computed within the Model Microfield Method for the conditions of
stellar atmospheres and circumstellar envelopes (\cite{Steh10}), and
\cite{Trem09} have performed improved calculations for the Stark
broadening of hydrogen lines in dense plasmas typical of white-dwarf
atmospheres.  The central asymmetry of the H$\beta$ line has been
measured and analysed (\cite{Djur09}) and new experimental results for
H$\alpha$ and H$\gamma$ have been published (\cite{Mija10a,Mija10b}).

Omar (2010, 2011) published new calculations for the 
Stark broadening of the He I lines at 504.8 nm, 388.9 nm, 318.8 nm,
667.8 nm and 501.6 nm formed in a dense plasma.
Tables of Stark broadening for the He I 447.1 nm line have been
generated using computer simulations (\cite{Gigo09}).  This line and
its forbidden component have also been studied theoretically
(\cite{Gonz11}) and experimentally (\cite{Ivko10,Gonz11}).
\cite{Gaoh08} have carried out experiments for the He I 388.9 nm and
706.5 nm lines.

\section{Broadening by neutral atoms and molecules}

The analysis of experimental molecular spectra in order to extract line
shape parameters is often very difficult.  Line shapes can be affected
by collisional narrowing and the dependence of collisional broadening
and shifting on molecular speed.  When these effects are sufficiently
important, fitting Voigt profiles to experimental spectra produces
systematic errors in the parameters retrieved.  Here the experimental
and theoretical results selected have been confined to the basic
atomic and molecular data required for a description of the pressure
broadening and shift of lines and molecular bands. 

   Since the last report an important book has been published
(\cite{Hart08}) that gives a comprehensive review of experimental and
theoretical work on collisional effects in molecular spectra.
In the following sections the items are labelled by `E' and `T' to
indicate experimental work and theoretical analysis, respectively.

\subsection{Broadening and shift of atomic lines}

New research has been published in the period 2008-2011 and the
transitions studied together with the perturbing atoms or molecules
are listed below.  The work is theoretical except where indicated by 'E'.
\\[1ex]
H: line wings of Ly$\alpha$ broadened by H and He (\cite{Alla09a},
 \cite{Alla09}); line wings of Ly$\gamma$ by H$^+$ (\cite{Alla09b})
and H$\alpha$ by H (\cite{Alla08}).\\
He: self broadening of line 3s$^3$S--2p$^3$P (\cite{Alla09c,Alla11}).\\
Li: self broadening of resonance line (\cite{Regg09}); resonance line
broadened by He (\cite{Peac09,Peac10a,Peac10b});
2s-3d transition broadened by Ne and Ar (\cite{Rose11}).\\
Na: resonance line broadened by H (\cite{Peac10b}); lines 3s-3p and
3p-3d broadened by He (\cite{Peac09,Peac10a,Peac10b}).\\
K: self broadening of resonance line (\cite{Regg09}) and line wings
(\cite{Talb08}).\\
K, Rb and Cs: self broadening of principal series (E) (\cite{Vadl09}).\\
Rb: 5s-5p D2 line, broadening by He, CH$_4$, C$_2$H$_6$,
C$_3$H$_8$, n-C$_4$H$_{10}$ (E) (\cite{Zame11}).\\
Cs: 6s-6p D2 line, broadening by $^3$He, H$_2$, HD, D$_2$,
N$_2$, CH$_4$, C$_2$H$_6$, CF$_4$ (E) (\cite{Pitz10}).\\

\subsection{Broadening and shift of molecular lines}

Much new data have been published since the last report was prepared.
The molecules are listed below with their perturbing atomic or
molecular species and are labelled by `E' and `T' to indicate
experimental work and theoretical analysis, respectively.
\\[1ex]
H$_2$-Ar: collision-induced absorption (T) (\cite{Tran11b}).\\
D$_2$-Kr: collision-induced absorption (E) (\cite{Abuk10}).\\
HI: lines broadened by N$_2$ (E) (\cite{Doma11}).\\
HBr: self broadening (E) (\cite{Doma09}).\\
HI and HBr: lines broadened by rare gases (E) (\cite{Doma09}).\\
HDO: lines broadened by CO$_2$ (T) (\cite{Gama11}).\\
HCl: lines broadened and shifted by N$_2$, He, Ar and Xe (E)
(\cite{Hurt09}).\\
HCN: lines broadened by N$_2$, O$_2$ and air (E)
(\cite{Yang08}).\\
H$_2$CO: lines broadened by H$_2$CO and N$_2$ (E+T)
(\cite{Jacq10}).\\
HNO$_3$: lines broadened by N$_2$ (T) (\cite{Lara09}).\\
HO$_2$: lines broadened by N$_2$ (E) (\cite{Miya11}).\\
H$_2$O$_2$: lines broadened by N$_2$, O$_2$ and air (E)
(\cite{Sato10}).\\
H$_2$O: lines broadened by H$_2$ (E) (\cite{Krup10}), (T)
(\cite{Wies10}); by H$_2$ and He (E) (\cite{Dick10});
 by N$_2$ (E) (\cite{Lavr10}; by N$_2$ and
O$_2$ (T) (\cite{Gama09}); by O$_2$ (E) (\cite{Petr11}; by
H$_2$O (E) (\cite{Lisa09,Ptas10}); by H$_2$O, N$_2$, O$_2$ (E+T)
(\cite{Cazz08,Cazz09,Kosh11}); by air (T) (\cite{Voro10});  
by CO$_2$ (T) (\cite{Saga09}); by H$_2$, He, N$_2$, O$_2$ and
CO$_2$ (E) (\cite{Dick09b}); by rare gases (E+T) (\cite{Fiad08}).\\
CH$_4$: lines broadened by N$_2$ (T) (\cite{Gaba10}); by N$_2$
and O$_2$ (E) (\cite{Lyul09}); by CH$_4$ (E)
(\cite{Smit10,Lyul11}); by CH$_4$ and N$_2$ (E) (\cite{Mcra11});
by O$_2$ and air (E) (\cite{Mart09}); by air (E) (\cite{Smit09,Smit11}).\\
C$_2$H$_2$: broadened by H$_2$ (T) (\cite{Thib11a}); by H$_2$ and
D$_2$ (E+T) (\cite{Thib09}); by N$_2$ (E) (\cite{Dhyn09,Fiss09,Dhyn10});
by C$_2$H$_2$ (E) (\cite{Lijs10,Pove11,Dhyn11});
by He and Ar (T) (\cite{Ivan10}); by Ne and Kr (E) (\cite{Nguy09a}).\\
C$_2$H$_4$: lines broadened by C$_2$H$_4$ (E) (\cite{Flau11});
by Ar (E+T) (\cite{Nguy09b}).\\
C$_2$H$_6$: lines broadened by N$_2$ (E) (\cite{Blan09}); by
C$_2$H$_6$ and N$_2$ (E) (\cite{Devi10b,Devi10c}); by O$_2$ and
air (E) (\cite{Fiss10}).\\
CH$_3$Br: lines broadened by N$_2$ (T) (\cite{Bous11}); by CH$_3$Br,
N$_2$ and O$_2$ (E) (\cite{Hoff09}); by CH$_3$Br (T) (\cite{Gome10}).\\
CH$_3$F: lines broadened by CH$_3$F and He (E) (\cite{Koub11}).\\
CO: lines broadened by CO, N$_2$ and O$_2$ (E) (\cite{Kosh09}); by
H$_2$, N$_2$, O$_2$, CO, CO$_2$ and He (E) (\cite{Dick09a}).\\
CO$_2$: by O$_2$ (E) (\cite{Devi10a}); by CO$_2$ (E+T)
(\cite{Pred10,Tran11a}); by CO$_2$, N$_2$ and O$_2$ (E) (\cite{Lijs08});
by air (T) (\cite{Hart09}), (E) (\cite{Guli10}), (E+T) (\cite{Lamo10});
by He (E) (\cite{Deng09}); by air and Ar (E) (\cite{Faro10}).\\
Cs$_2$: lines broadened by Cs$_2$ (E) (\cite{Misa09}).\\
N$_2$: lines broadened by H$_2$ (T) (\cite{Gome11}); by N$_2$ (E+T)
(\cite{Thib11b}).\\
NH$_3$: lines broadened by H$_2$ and He (E) (\cite{Hanl09}); by He (T)
(\cite{Dhib10}; by NH$_3$ (E) (\cite{Arou09,Guin11}); by NH$_3$ and
O$_2$ (E+T) (\cite{Nour09}).\\
O$_2$: lines broadened by O$_2$ (E) (\cite{Lisa10,Wojt11});
by O$_2$ and OO isotopologues (E) (\cite{Long11}); by O$_2$ and air
(E) (\cite{Long10}).\\
O$_2$-CO$_2$: collision-induced absorption (E) (\cite{Vang09}).\\
O$_3$: lines broadened by air (\cite{Drou08}); by N$_2$ and air (E+T)
  (\cite{Tran11c}).\\
OH: lines broadened by N$_2$, H$_2$O and Ar (E) (\cite{Hwan08}).\\
OCS lines broadened by N$_2$, O$_2$ and OCS (E) (\cite{Kose09});
by N$_2$ and O$_2$ (E) (\cite{Gala11}).\\ 
I$_2$: lines broadened by Ar (E) (\cite{Phil08}).\\

\section[Databases]{Databases}

Some useful databases are:

Vienna Atomic Line Database (VALD) of atomic data for analysis of
radiation from astrophysical objects, containing central wavelengths,
energy levels, statistical weights, transition probabilities and line
broadening parameters for all chemical elements of astronomical importance.
It can be found at http://vald.astro.univie.ac.at/ (\cite{Kupk99}).

The database of Robert L. Kurucz comprises atomic line parameters, including
line broadening. An update to this database is discussed by \cite{Kur11}. 
\linebreak
(http://kurucz.harvard.edu)

CHIANTI database (\cite{Dere09}) contains a critically evaluated set of
up-to-
\linebreak
date atomic data for the analysis of optically thin collisionally
ionized astrophysical plasmas.  It lists experimental and calculated 
wavelengths, radiative data and rates for electron and proton collisions,
see websites http://sohowww.nascom.nasa.gov/solarsoft and
http://www.damtp.cam.ac.uk/user/astro/chianti/.

CDMS -- Cologne Database for Molecular Spectroscopy, see website\\ 
http://www.ph1.uni-koeln.de/vorhersagen/, provides recommendations for
spectroscopic transition frequencies and intensities for atoms and
molecules of astronomical interest in the frequency range 0-10 THz, i.e.
0-340 cm$^{-1}$ (\cite{Mull05}).

BASECOL database (http://basecol.obspm.fr) contains excitation rate
coefficients for ro-vibrational excitation of molecules by electrons,
He and H$_2$ and it is mainly used for the study of interstellar,
circumstellar and cometary atmospheres.

TIPTOPbase (http://cdsweb.u-strasbg.fr/topbase/home.html) contains:\\
 (i) TOPbase, that lists atomic data computed in the Opacity Project;
namely LS-coupling energy levels, gf-values and photoionization cross
sections for light elements (Z $\le$ 26) of astrophysical interest and \\
(ii) TIPbase that lists intermediate-coupling energy levels, transition probabilities and 
electron impact excitation cross sections and rates for astrophysical
applications (Z $\le$ 28), computed by the IRON Project.

HITRAN -- (HIgh-resolution TRANsmission molecular absorption database) 
is at
\linebreak
http://www.cfa. harvard.edu/hitran/ (\cite{Roth09}). It lists
individual line parameters for molecules in the gas phase (microwave
through to the UV), photoabsorption cross-sections for many molecules,
and refractive indices of several atmospheric aerosols.  A high
temperature extension to HITRAN is HITEMP (To access the HITEMP data:
ftp to
cfa-ftp.harvard.edu; user = anonymous; password = e-mail address).
It contains data for water, CO$_2$, CO, NO and OH (\cite{Roth10}).

GEISA -- (Gestion et Etude des Informations Spectroscopiques
Atmosph\'eriques) is a computer-accessible spectroscopic database, 
designed to facilitate accurate forward radiative transfer calculations
using a line-by-line and layer-by-layer approach. It can be found at
http://ether.ipsl.jussieu.fr/etherTypo/?id=950 (\cite{Jacq08}).

NIST -- The National Institute of Standards and Technology hosts a
number of useful databases for Atomic and Molecular Physics.
A list can be found at\\
http://www.nist.gov/srd/atomic.cfm.  Among them are: An atomic spectra
database and three bibliographic databases providing references on
atomic energy levels and spectra, transition probabilities and spectral
line shapes and line broadening.

STARK-B database (http://stark-b.obspm.fr) contains theoretical widths 
and shifts of isolated lines of atoms and ions due to collisions with
charged perturbers, obtained using the impact approximation
(\cite{Saha10}).

The European FP7 project will finish at the end of 2012.  The virtual
Atomic and Molecular Data Centre (VAMDC - http://www.vamdc.eu/) is being
created with the aim of building an accessible and interoperable
e-infrastructure for atomic and molecular data that will upgrade and 
integrate European (and other) A\&M database services (\cite{Dube11},
\cite{Rixo11}).

\end{document}